\documentclass[3p,fleqn,a4paper]{elsarticle}

\usepackage[T1]{fontenc}
\usepackage{amsmath,amssymb,amsfonts,mathtools}
\usepackage{bm}
\usepackage{graphicx}
\usepackage{subcaption}
\usepackage{booktabs}
\usepackage{siunitx}
\usepackage{xcolor}
\usepackage{hyperref}
\usepackage{enumitem}
\usepackage{multirow}
\usepackage{array}
\usepackage{makecell}
\usepackage{placeins}
\usepackage{float}
\usepackage[mathlines]{lineno}
\usepackage{microtype}

\hypersetup{colorlinks=true,linkcolor=black,citecolor=blue,urlcolor=blue}
\graphicspath{{figures/}{./}}
\newcommand{\bX}{\bm{X}}
\newcommand{\bx}{\bm{x}}
\newcommand{\bu}{\bm{u}}

\newcommand{\bP}{\bm{P}}
\newcommand{\bsig}{\bm{\sigma}}
\newcommand{\bF}{\bm{F}}

\newcommand{\bI}{\bm{I}}
\newcommand{\bN}{\bm{N}}
\newcommand{\bp}{\bm{p}}
\newcommand{\bZ}{\bm{Z}}
\newcommand{\R}{\mathbb{R}}
\newcommand{\Div}{\operatorname{Div}}
\newcommand{\tr}{\operatorname{tr}}

\newcommand{\cL}{\mathcal{L}}

\newcommand{\cD}{\mathcal{D}}
\newcommand{\cB}{\mathcal{B}}
\newcommand{\cQ}{\mathcal{Q}}
\newcommand{\cS}{\mathcal{S}}
\newcommand{\todofig}[1]{\begin{center}\fbox{\begin{minipage}{0.88\linewidth}\centering\vspace{0.8cm}\textcolor{blue}{#1}\vspace{0.8cm}\end{minipage}}\end{center}}
\newcommand{\maybeincludegraphics}[2][]{\IfFileExists{#2}{\includegraphics[#1]{#2}}{\todofig{Missing figure file: \texttt{\detokenize{#2}}}}}

\journal{Computer Methods in Applied Mechanics and Engineering}

\begin{document}
	
	
	\begin{frontmatter}
		
		\title{PI-GINOT: Data-free geometry-informed neural operator learning for finite-strain hyperelasticity on parametric DogBone specimens}
		
		\author[SUST]{Aamir Dean\corref{cor1}}
		\ead{a.dean@sustech.edu}
		\cortext[cor1]{Corresponding author}
		
		\author[SUST]{Betim Bahtiri}
		
		\address[SUST]{School of Civil Engineering, College of Engineering, Sudan University of Science and Technology, P.O. Box 72, Khartoum, Sudan}
		
		\begin{abstract}
			Parametric nonlinear solid-mechanics simulations are central to virtual testing, structural optimisation and uncertainty-aware design, yet repeated high-fidelity finite-element analyses remain expensive when geometry varies. Physics-informed neural networks reduce the dependence on labelled simulation data, but conventional formulations usually approximate one boundary-value problem at a time and must be retrained when the domain changes. This paper introduces PI-GINOT, a physics-informed geometry-informed neural operator transformer for data-free finite-strain hyperelasticity on a controlled four-parameter family of DogBone tensile specimens. The study is deliberately restricted to a fixed material model, fixed displacement-controlled loading and a smooth single-topology geometry family, so that the central question--whether a reusable geometry-conditioned operator can be trained without finite-element field labels--can be assessed transparently. Each geometry is represented by a boundary point cloud that is encoded into latent geometry tokens; a cross-attention physics decoder then predicts displacement fields at arbitrary query points. Essential displacement boundary conditions are imposed exactly through a hard boundary-condition layer, while stresses are obtained by automatic differentiation and a compressible Neo-Hookean plane-stress constitutive model. The training objective combines reference-configuration equilibrium, traction-free boundary residuals, symmetry traction components, a determinant barrier and an internal axial section-force consistency term. {\color{black}Independent Abaqus/Standard simulations using a custom matching user element are performed only after training and serve exclusively as finite-element reference data.} Across eight independent comparison geometries, the final PI-GINOT model predicts displacement fields with relative $L^2$ errors between $2.1\%$ and $7.1\%$, peak von Mises stresses within $0.9\%$--$13.3\%$, and internal section-force mean absolute errors not exceeding $10.3\%$. The remaining errors are concentrated in component-wise stress fields, which range from $10.0\%$ to $47.6\%$ and are largest for narrow-gauge specimens, identifying local stress-gradient resolution near the gauge-fillet transition as the main limitation. The results demonstrate a bounded but mechanically meaningful data-free geometry-conditioned operator for nonlinear solid mechanics and provide a transparent validation protocol for future neural-operator developments in computational mechanics.
		\end{abstract}
		
		
	\end{frontmatter}
	
	\section*{Nomenclature}
	\noindent
	\begin{tabular}{p{2.4cm}p{10.2cm}}
		$\Omega(\bp)$ & Reference domain parameterised by geometry vector $\bp$ \\
		$\partial\Omega$ & Boundary of the reference domain \\
		$\Gamma_u,\Gamma_t$ & Essential-displacement and traction boundaries \\
		$\bX=(X,Y)$ & Material/reference coordinates \\
		$\bu=(u,v)$ & In-plane displacement field \\
		$\bar u$ & Prescribed axial displacement at the grip boundary \\
		$\bF$ & Deformation gradient, $\bF=\bI+\nabla_{\bX}\bu$ \\
		$J$ & Determinant of the deformation gradient \\
		$\Psi$ & Compressible Neo-Hookean strain-energy density \\
		$\bP$ & First Piola--Kirchhoff stress tensor \\
		$\bsig$ & Cauchy stress tensor used for post-processing \\
		$E,\nu$ & Young's modulus and Poisson's ratio \\
		$\lambda,\mu$ & Lam\'e parameters \\
		$\bp$ & Geometry vector $[L_{\rm total},W_{\rm grip},W_{\rm gauge},R_{\rm fillet}]^T$ \\
		$\bZ$ & Latent geometry-token matrix generated by the geometry encoder \\
		$\phi_u,\phi_v$ & Raw decoder correction fields before hard boundary-condition enforcement \\
		$N(s)$ & Internal axial section force through the vertical section $X=s$ \\
		$\cL_{\rm eq}$ & Interior equilibrium residual loss \\
		$\cL_{\rm trac}$ & Traction-free boundary residual loss \\
		$\cL_{\rm part}$ & Partial traction/symmetry residual loss \\
		$\cL_{\rm bar}$ & Determinant-barrier loss \\
		$\cL_N$ & Axial section-force consistency loss \\
	\end{tabular}

	\section{Introduction}
	\label{sec:introduction}
	
	High-fidelity computational solid mechanics is increasingly used to explore families of designs rather than to analyse a single structure in isolation. Virtual testing, reliability assessment, shape optimisation, uncertainty quantification and digital-twin workflows all require repeated solution of related boundary-value problems. In nonlinear mechanics, this repeated-query setting remains expensive because each new geometry or load case may require geometry construction, meshing, nonlinear iterations, convergence checks and field post-processing. The cost is especially restrictive when the required output is a full displacement and stress field, not only a scalar force or stiffness measure. Full-field quantities are central to stress concentration assessment, fatigue initiation, fracture modelling and design optimisation, and therefore cannot easily be replaced by low-dimensional response surfaces.
	
	The finite-element method remains the standard tool for reliable nonlinear solid mechanics because it provides a systematic variational framework, mature element technology and well-tested nonlinear solvers \citep{Simo1998Computational,Bonet2008Nonlinear,Holzapfel2000Nonlinear}. However, conventional finite-element analysis is not designed to provide immediate responses over a high-dimensional design space. Reduced-order models and surrogate models can accelerate repeated analyses, but they typically require labelled high-fidelity snapshots and often rely on a fixed mesh, a fixed topology or a carefully constructed reduced basis. These assumptions are restrictive when the geometry itself is the parameter of interest. As the geometry changes, boundary normals, stress concentrations, admissible boundary conditions and the required sampling density change as well. A useful mechanics surrogate for such settings must therefore be geometry aware, reusable over a family of domains and constrained by the governing physics.
	
	Physics-informed neural networks (PINNs) introduced a mesh-free differentiable route for solving forward and inverse problems by training neural fields with residuals of the governing equations and boundary conditions \citep{Raissi2019,Karniadakis2021PINN}. This idea is attractive in computational mechanics because balance laws, kinematics, constitutive equations and boundary tractions are known a priori. In solid mechanics, physics-informed collocation and energy-based neural methods have been applied to elasticity, finite deformation, elastodynamics and inverse identification \citep{Samaniego2020DEM,Nguyen2020DEM,Haghighat2020Solid,Rao2020Elastodynamics,Abueidda2021Energy,Haghighat2021SciANN}. Exact or constructive enforcement of essential boundary conditions has also been recognised as important, because soft boundary penalties may compete with equilibrium terms and can contaminate stresses and reactions \citep{Sukumar2022PINN}. Despite these strengths, a standard PINN is usually an instance-specific solver: after training on one geometry and one set of boundary conditions, it is not automatically a reusable surrogate for another specimen.
	
	Neural operators address this limitation by learning maps between input and output functions rather than one solution field at a time. DeepONet \citep{Lu2021DeepONet}, graph and multipole graph neural operators \citep{Li2020GKN,Li2020MGNO}, Fourier neural operators \citep{Li2021FNO} and the general neural-operator framework \citep{Kovachki2023NeuralOperator} established the operator-learning viewpoint for parametric partial differential equations. Once trained, an operator can be evaluated for new inputs without retraining, which matches the multi-query needs of design and uncertainty analysis. Nevertheless, several successful neural operators are most natural on fixed grids or fixed computational domains. This is a serious restriction for solid mechanics, where the domain shape and boundary partition are often the dominant design variables.
	
	Recent geometry-aware operator models aim to remove this fixed-domain restriction by representing the geometry using point clouds, meshes, graphs, signed-distance fields or boundary descriptors. Physics-Informed Geometry-Aware Neural Operator (PI-GANO) combines geometry encoding and physics-informed training for variable-domain PDE problems \citep{Zhong2025PIGANO}. Point Cloud Neural Operator (PCNO) develops operator learning directly on point clouds for irregular domains \citep{Zeng2025PCNO}. Geometry-Informed Neural Operator Transformer (GINOT) is particularly relevant because it encodes surface point clouds into latent geometry tokens and conditions a query-point decoder through attention mechanisms \citep{Liu2025GINOT}. This architecture is natural for mechanics: a boundary point cloud describes the specimen shape, while a decoder can be queried at arbitrary interior, boundary or post-processing points. However, finite-strain solid mechanics adds requirements beyond generic PDE interpolation. The displacement field must generate a physically admissible deformation gradient, the determinant of the deformation gradient must remain positive, stresses must be computed consistently from a hyperelastic strain-energy density, and displacement boundary conditions must be satisfied accurately.
	
	The present work develops PI-GINOT, a physics-informed geometry-informed neural operator transformer for data-free finite-strain hyperelasticity on a parametric DogBone tensile specimen. The benchmark is deliberately controlled but mechanically meaningful. The purpose is not to claim geometry universality over arbitrary solid domains, but to establish a reproducible mechanics benchmark in which geometry variation, exact boundary-condition enforcement, physics-informed training and strict post-training finite-element comparison can be examined without confounding effects from topology changes, material heterogeneity or load-path variation. The specimen family has a simple topology, yet includes key difficulties for geometry-conditioned mechanics: variable specimen length, grip width, gauge width and fillet radius; curved traction-free boundaries; stress concentration near the gauge-fillet transition; symmetry constraints; and a displacement-controlled grip. The material response is formulated as compressible Neo-Hookean hyperelasticity in the reference configuration, with a plane-stress closure enforced at the constitutive level. The neural operator is trained without finite-element displacement or stress labels. Its objective combines local equilibrium, traction-free boundary conditions, symmetry-related partial traction residuals, determinant regularisation and an internal axial section-force consistency term.
	
	The central methodological idea is to separate geometry encoding from physics decoding. Each specimen boundary is encoded once into latent geometry tokens. The decoder is then evaluated repeatedly at arbitrary query points to assemble interior residuals, boundary residuals, section-force integrals or validation fields. Essential displacement boundary conditions are imposed exactly through a hard boundary-condition layer, so the network does not learn symmetry and grip constraints through soft penalties. Stresses are not independently fitted; they are obtained by automatic differentiation of the admissible displacement field and by the hyperelastic constitutive law. Independent strict finite-element simulations, implemented with a matching Abaqus user element, are used only after training for validation.
	
	The contributions of this work are:
	\begin{enumerate}[leftmargin=*,itemsep=2pt]
		\item a data-free, geometry-conditioned neural operator for finite-strain hyperelastic solid mechanics on a four-parameter family of DogBone geometries;
		\item a PI-GINOT architecture combining boundary-point-cloud geometry encoding, cross-attention physics decoding and exact hard enforcement of essential displacement boundary conditions;
		\item a reference-configuration physics objective combining local equilibrium, traction-free boundaries, symmetry-related partial traction residuals, determinant-barrier regularisation and internal axial section-force consistency;
		\item a strict independent finite-element validation protocol using a matching Abaqus user element for the same geometry, material law, plane-stress closure and displacement loading;
		\item a mechanics-oriented assessment over eight independent post-training finite-element comparison cases, including displacement errors, stress errors, peak von Mises errors and internal load-transfer consistency.
	\end{enumerate}
	
	The paper is organised as follows. Section~\ref{sec:problem} defines the parametric DogBone boundary-value problem. Section~\ref{sec:mechanics} presents the finite-strain hyperelastic formulation and plane-stress closure. Section~\ref{sec:architecture} introduces the PI-GINOT architecture and hard boundary-condition layer. Section~\ref{sec:loss} derives the physics-informed objective. Section~\ref{sec:training} describes the training configuration and diagnostics. Section~\ref{sec:fem} explains the strict finite-element validation protocol. Section~\ref{sec:results} presents the validation results. Section~\ref{sec:discussion} discusses interpretation, limitations and extensions, and Section~\ref{sec:conclusion} concludes the paper.
	
	\section{Parametric DogBone boundary-value problem}
	\label{sec:problem}
	
	\subsection{Geometry and quarter-domain reduction}
	
	Let $\Omega(\bp)\subset\R^2$ denote the reference domain of a DogBone tensile specimen parameterised by
	\begin{equation}
		\bp=\left[L_{\rm total},\,W_{\rm grip},\,W_{\rm gauge},\,R_{\rm fillet}\right]^T .
	\end{equation}
	The computational model uses the upper-right quarter of the full specimen, exploiting symmetry about the longitudinal and transverse centre lines. Thus
	\begin{equation}
		0\le X\le L_{\rm half}=\frac{L_{\rm total}}{2},
		\qquad
		0\le Y\le H_{\rm grip}=\frac{W_{\rm grip}}{2},
		\qquad
		H_{\rm gauge}=\frac{W_{\rm gauge}}{2} .
	\end{equation}
	A circular fillet connects the gauge region to the grip region. The sampled geometry range is
	\begin{equation}
		\begin{split}
			L_{\rm total}&\in[40,70]~{\rm mm},\qquad
			W_{\rm grip}\in[16,26]~{\rm mm},\\
			W_{\rm gauge}&\in[6,14]~{\rm mm},\qquad
			R_{\rm fillet}\in[8,20]~{\rm mm}.
		\end{split}
		\label{eq:geometry_ranges}
	\end{equation}
	The training implementation generates boundary, interior and section-integration points analytically. No finite-element mesh is used during PI-GINOT training.
	
	\subsection{Boundary conditions and loading}
	
	The essential displacement conditions are imposed on the symmetry and grip boundaries. On the vertical symmetry boundary $X=0$, the axial displacement is zero; on the horizontal symmetry boundary $Y=0$, the transverse displacement is zero. On the right grip boundary $X=L_{\rm half}$, a prescribed axial displacement $\bar u$ is applied. In the present study,
	\begin{equation}
		\bar u=1~{\rm mm}.
	\end{equation}
	The essential conditions are therefore
	\begin{equation}
		u(0,Y)=0,
		\qquad
		u(L_{\rm half},Y)=\bar u,
		\qquad
		v(X,0)=0 .
		\label{eq:essential_bc}
	\end{equation}
	The remaining external boundaries of the quarter model are traction-free. In the reference configuration the traction condition is
	\begin{equation}
		\bP\bN=\bm{0}\qquad \text{on }\Gamma_t,
		\label{eq:traction_free}
	\end{equation}
	where $\bN$ is the outward unit normal in the reference configuration. On the symmetry planes, the corresponding displacement component is already imposed by Eq.~\eqref{eq:essential_bc}; the remaining natural traction component is included in the physics loss.
	
\begin{figure}[t]
		\centering
		\maybeincludegraphics[width=0.99\linewidth]{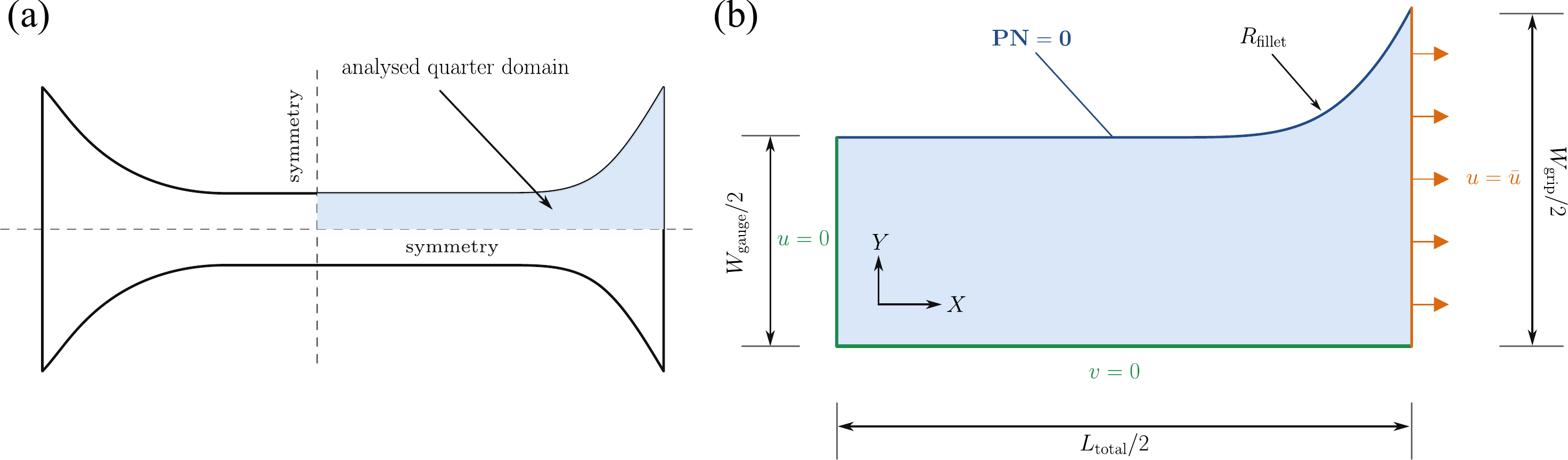}
		\caption{DogBone geometry and analysed quarter-domain boundary-value problem. Panel (a) shows the full symmetric specimen and the analysed quarter domain. Panel (b) shows the quarter-domain model with symmetry conditions on the left and bottom boundaries, a prescribed axial displacement $u=\bar{u}$ on the right grip edge and traction-free boundary conditions on the top and fillet boundary.	The geometric parameters are the total specimen length $L_{\mathrm{total}}$, grip width $W_{\mathrm{grip}}$, gauge width $W_{\mathrm{gauge}}$ and fillet radius $R_{\mathrm{fillet}}$.}
		\label{fig:problem_geometry}
	\end{figure}

	The resulting full-specimen geometry, analysed quarter domain and boundary-condition assignment are shown in Fig.~\ref{fig:problem_geometry}.
	
	\subsection{Operator-learning formulation}
	
	The goal is to learn one geometry-conditioned solution operator rather than a separate neural approximation for each specimen. Formally, the operator maps the boundary representation, loading and query coordinate to the mechanical response:
	\begin{equation}
		\cS:\left(\partial\Omega(\bp),\bar u,\bX\right)
		\mapsto
		\left(\bu(\bX;\bp,\bar u),\bP(\bX;\bp,\bar u),\bsig(\bX;\bp,\bar u)\right).
		\label{eq:operator_map}
	\end{equation}
	In PI-GINOT, $\partial\Omega(\bp)$ is represented by a boundary point cloud. The network directly predicts an admissible displacement field; stresses are then computed by automatic differentiation of the displacement field and the hyperelastic constitutive law. This design allows the same encoded geometry to be reused for interior residuals, boundary residuals, section-force integration and post-processing queries. Although $\bar u$ is included in Eq.~\eqref{eq:operator_map} to make the prescribed loading explicit, it is fixed to $1~\mathrm{mm}$ throughout the present study and is not used as an additional operator input dimension.
	
	\section{Finite-strain hyperelastic formulation}
	\label{sec:mechanics}
	
	\subsection{Kinematics}
	
	All governing equations are written in the reference configuration. The deformation map is
	\begin{equation}
		\bx(\bX)=\bX+\bu(\bX),
	\end{equation}
	and the deformation gradient is
	\begin{equation}
		\bF=\frac{\partial \bx}{\partial\bX}=\bI+\nabla_{\bX}\bu .
		\label{eq:defgrad}
	\end{equation}
	For the in-plane displacement field $\bu=(u,v)$,
	\begin{equation}
		\bF_{2D}=
		\begin{bmatrix}
			1+u_{,X} & u_{,Y}\\
			v_{,X} & 1+v_{,Y}
		\end{bmatrix},
		\qquad
		J_{2D}=\det\bF_{2D} .
	\end{equation}
	Because the material model is formulated in three dimensions while the specimen is treated as plane stress, the full deformation gradient is represented as
	\begin{equation}
		\bF=\begin{bmatrix}
			\bF_{2D} & \bm{0}\\
			\bm{0}^T & F_{33}
		\end{bmatrix},
		\qquad
		J=J_{2D}F_{33} .
		\label{eq:F3d_plane_stress}
	\end{equation}
	
	\subsection{Compressible Neo-Hookean material and plane-stress closure}
	
	The strain-energy density is taken as the logarithmic compressible Neo-Hookean form
	\begin{equation}
		\Psi(\bF)=\frac{\mu}{2}\left(\tr(\bF^T\bF)-3\right)-\mu\ln J+\frac{\lambda}{2}(\ln J)^2,
		\label{eq:neo_hookean_energy}
	\end{equation}
	with Lam\'e parameters
	\begin{equation}
		\mu=\frac{E}{2(1+\nu)},
		\qquad
		\lambda=\frac{E\nu}{(1+\nu)(1-2\nu)}.
	\end{equation}
	The material parameters are
	\begin{equation}
		E=760~{\rm MPa},
		\qquad
		\nu=0.23,
	\end{equation}
	so that $\mu=308.94~{\rm MPa}$ and $\lambda=263.16~{\rm MPa}$. The first Piola--Kirchhoff stress is
	\begin{equation}
		\bP=\frac{\partial \Psi}{\partial\bF}
		=\mu\bF+\left(\lambda\ln J-\mu\right)\bF^{-T}.
		\label{eq:piola_stress}
	\end{equation}
	Plane stress is enforced by solving the local scalar closure
	\begin{equation}
		P_{33}(\bF_{2D},F_{33})=0
		\label{eq:plane_stress_closure}
	\end{equation}
	for $F_{33}$ at each query point. In the implementation this scalar equation is solved by a differentiable Newton iteration. The in-plane stress response then follows from Eqs.~\eqref{eq:F3d_plane_stress}--\eqref{eq:piola_stress}. For visualisation, the Cauchy stress is computed as
	\begin{equation}
		\bsig=\frac{1}{J}\bP\bF^T .
	\end{equation}
	
	\subsection{Strong form}
	
	In the absence of body forces, equilibrium in the reference configuration is
	\begin{equation}
		\Div\bP=\bm{0}\qquad\text{in }\Omega(\bp),
		\label{eq:strong_equilibrium}
	\end{equation}
	with boundary conditions given by Eqs.~\eqref{eq:essential_bc} and \eqref{eq:traction_free}. This reference-frame formulation is convenient for physics-informed training because all residuals are evaluated on the fixed reference geometry.
	
	\section{PI-GINOT architecture}
	\label{sec:architecture}
	
\begin{figure}[t]
		\centering
		\maybeincludegraphics[width=0.99\linewidth]{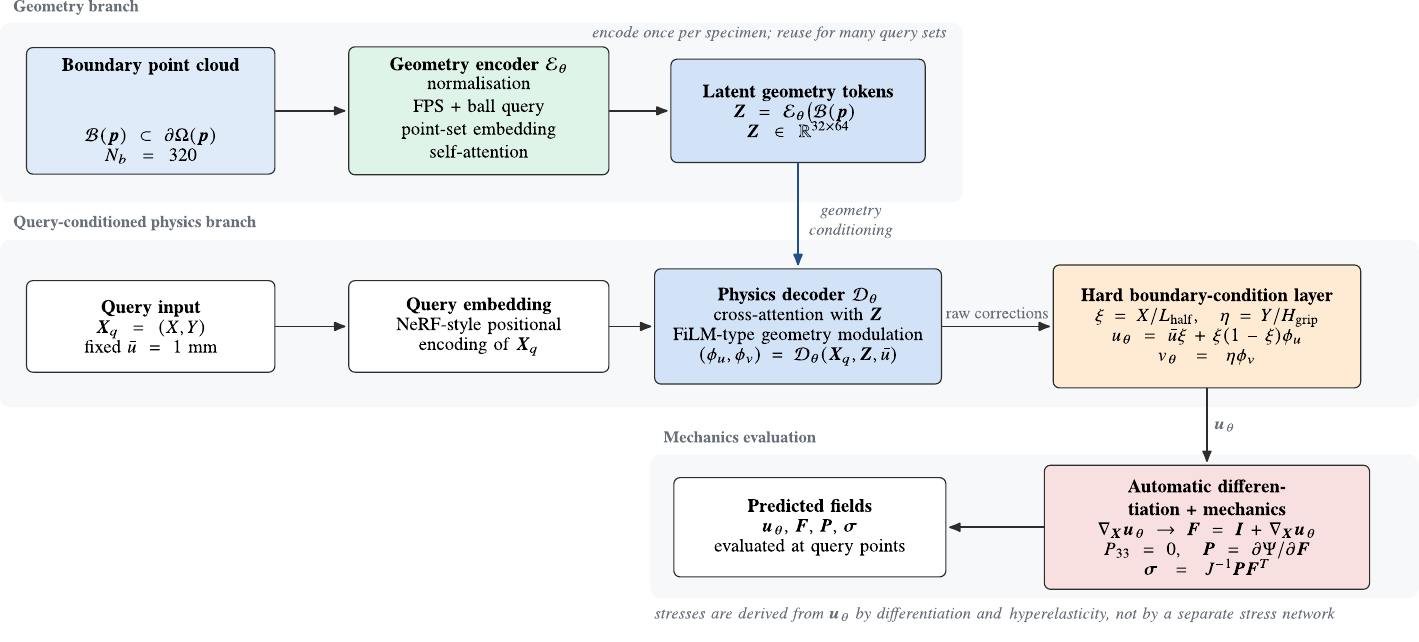}
		\caption{PI-GINOT architecture. A boundary point cloud sampled from $\partial\Omega(\bp)$ is encoded into latent geometry tokens $\bZ$. Query coordinates $\bX_q$ interact with these tokens through cross-attention and FiLM-type modulation in the physics decoder. The decoder returns raw correction fields $(\phi_u,\phi_v)$ that are mapped by a hard boundary-condition layer to an admissible displacement field. Stresses are obtained by automatic differentiation and the constitutive model rather than by a separate stress network.}
		\label{fig:architecture}
	\end{figure}

	PI-GINOT combines a boundary point-cloud geometry encoder with a query-point physics decoder, as illustrated in Fig.~\ref{fig:architecture}. The architecture follows the branch--trunk logic of neural operators, but uses attention to condition local field predictions on the encoded geometry. The geometry is encoded once per specimen and reused for many query sets, which is essential because interior, boundary and section-force points are sampled separately during training.
	
	\subsection{Geometry encoder}
	
	The boundary point cloud is denoted by
	\begin{equation}
		\cB(\bp)=\{\bX_b^{(i)}\}_{i=1}^{N_b}\subset\partial\Omega(\bp),
	\end{equation}
	with $N_b=320$ points in the present implementation. The point cloud is first normalised to a canonical coordinate range to reduce scale variation between geometries. Farthest-point sampling and local ball-query grouping are then used to form local neighbourhoods, followed by point-set embedding layers inspired by PointNet++ \citep{Qi2017PointNetPP}. Self-attention layers transform the local geometric features into a matrix of latent geometry tokens
	\begin{equation}
		\bZ=\mathcal{E}_\theta(\cB(\bp))\in\R^{N_{\rm tok}\times d},
		\label{eq:geometry_tokens}
	\end{equation}
	where $N_{\rm tok}=32$ and $d=64$ in the final model.
	
	\subsection{Physics decoder}
	
	The decoder receives query coordinates $\bX_q=(X,Y)$ and the geometry tokens $\bZ$. Query coordinates are embedded using a NeRF-style positional encoding \citep{Mildenhall2020NeRF}. The resulting query features interact with $\bZ$ through cross-attention blocks based on transformer attention \citep{Vaswani2017Attention}. FiLM-type modulation is applied using a pooled geometry representation so that the local field prediction depends on both the query location and the global specimen shape. The decoder returns raw correction fields
	\begin{equation}
		(\phi_u,\phi_v)=\cD_\theta(\bX_q,\bZ,\bar u).
	\end{equation}
	These raw fields are not interpreted directly as physical displacements. They are passed through the hard boundary-condition layer described next.
	
	\subsection{Hard enforcement of essential boundary conditions}
	
	Let
	\begin{equation}
		\xi=\frac{X}{L_{\rm half}},
		\qquad
		\eta=\frac{Y}{H_{\rm grip}}.
	\end{equation}
	The admissible displacement field is constructed as
	\begin{equation}
		\begin{split}
			u_\theta(X,Y)&=\bar u\,\xi+\xi(1-\xi)\,\phi_u(X,Y;\bZ),\\
			v_\theta(X,Y)&=\eta\,\phi_v(X,Y;\bZ).
		\end{split}
		\label{eq:hard_bc}
	\end{equation}
	Equation~\eqref{eq:hard_bc} enforces
	\begin{equation}
		u_\theta(0,Y)=0,
		\qquad
		u_\theta(L_{\rm half},Y)=\bar u,
		\qquad
		v_\theta(X,0)=0
	\end{equation}
	for any decoder output. Exact enforcement removes the need to balance soft displacement penalties against equilibrium and traction residuals and ensures that all training and validation predictions remain kinematically admissible with respect to the essential constraints.
	
	\section{Physics-informed loss formulation}
	\label{sec:loss}
	
	\subsection{Interior equilibrium loss}
	
	For interior collocation points $\cQ_{\rm int}\subset\Omega(\bp)$, equilibrium is enforced by
	\begin{equation}
		\cL_{\rm eq}=\frac{1}{|\cQ_{\rm int}|}\sum_{\bX_i\in\cQ_{\rm int}}
		\left\|\Div\bP(\bX_i)\right\|_2^2 .
		\label{eq:loss_eq}
	\end{equation}
	All derivatives are computed by automatic differentiation with respect to the reference coordinates.
	
	\subsection{Traction-free and symmetry traction losses}
	
	For traction-free boundary points $\cQ_t\subset\Gamma_t$,
	\begin{equation}
		\cL_{\rm trac}=\frac{1}{|\cQ_t|}\sum_{\bX_i\in\cQ_t}
		\left\|\bP(\bX_i)\bN(\bX_i)\right\|_2^2 .
		\label{eq:loss_trac}
	\end{equation}
	On the symmetry planes, the corresponding displacement component is imposed exactly. The complementary natural component is enforced through a partial traction loss. If $\bm{e}_\alpha$ denotes the traction component that should vanish on a symmetry segment $\Gamma_s$, then
	\begin{equation}
		\cL_{\rm part}=\frac{1}{|\cQ_s|}\sum_{\bX_i\in\cQ_s}
		\left(\bm{e}_\alpha\cdot\bP(\bX_i)\bN(\bX_i)\right)^2 .
		\label{eq:loss_part}
	\end{equation}
	
	\subsection{Determinant barrier}
	
	A determinant barrier is included to discourage locally inverted or nearly singular deformations. For a set of monitored points $\cQ_J$,
	\begin{equation}
		\cL_{\rm bar}=\frac{1}{|\cQ_J|}\sum_{\bX_i\in\cQ_J}
		\left\langle J_{\min}-J(\bX_i)\right\rangle_+^2,
		\label{eq:loss_barrier}
	\end{equation}
	where $\langle a\rangle_+=\max(a,0)$ and $J_{\min}$ is a small positive threshold.
	
	\subsection{Internal section-force consistency}
	
	Local residuals do not necessarily guarantee globally consistent load transfer on a finite collocation set. For a vertical section at $X=s$, define the internal axial force resultant
	\begin{equation}
		N(s)=2\int_{0}^{Y_{\rm top}(s)} P_{11}(s,Y)\,{\rm d}Y,
		\label{eq:section_force}
	\end{equation}
	where $Y_{\rm top}(s)$ is the upper boundary of the quarter-domain section and the factor two accounts for symmetry about the horizontal centre line. The constant factor does not affect the coefficient of variation, but it gives the full-height axial resultant corresponding to the half-specimen in the transverse direction. For section locations $\{s_k\}_{k=1}^{N_s}$,
	\begin{equation}
		\cL_N=\frac{1}{N_s}\sum_{k=1}^{N_s}
		\left(\frac{N(s_k)-\bar N}{|\bar N|+\epsilon_N}\right)^2,
		\qquad
		\bar N=\frac{1}{N_s}\sum_{k=1}^{N_s}N(s_k).
		\label{eq:loss_section_force}
	\end{equation}
	This term penalises variations of the axial resultant along the specimen and therefore constrains global equilibrium in a way that complements the pointwise divergence residual.
	
	\subsection{Composite objective}
	
	The final physics-informed objective is
	\begin{equation}
		\cL=
		w_{\rm eq}\cL_{\rm eq}
		+w_{\rm trac}\cL_{\rm trac}
		+w_{\rm part}\cL_{\rm part}
		+w_{\rm bar}\cL_{\rm bar}
		+w_N\cL_N .
		\label{eq:composite_loss}
	\end{equation}
	The relative weights compensate for the different natural scales of equilibrium, traction and section-force residuals. In the final configuration, additional emphasis is placed on curved traction-free boundaries and section-force consistency because these terms were found to be important for narrow-gauge geometries.
	
\begin{figure}[t]
		\centering
		\maybeincludegraphics[width=0.98\linewidth]{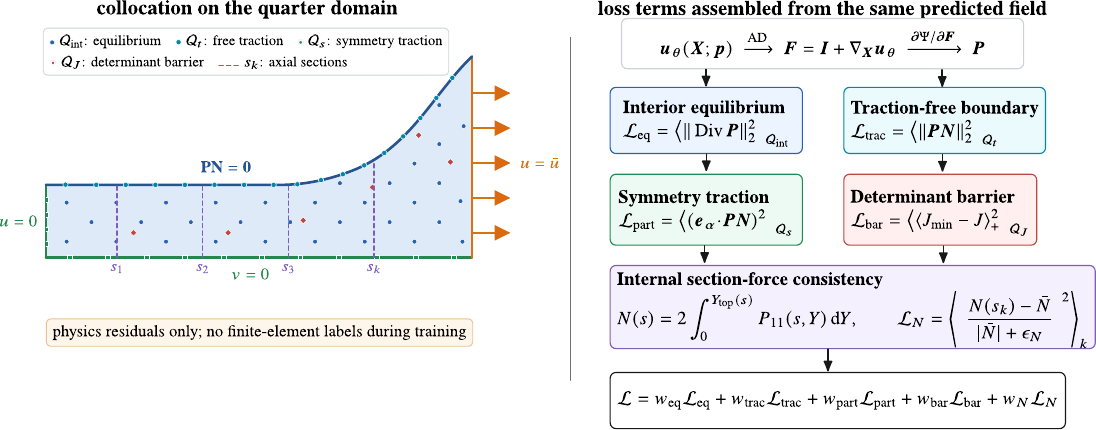}
		\caption{Physics-informed loss assembly. Interior collocation points enforce $\Div\bP=\bm{0}$, traction-free boundary points enforce $\bP\bN=\bm{0}$, symmetry boundaries enforce complementary traction components, and vertical sections provide a global axial section-force consistency check.}
		\label{fig:loss_assembly}
	\end{figure}

	The complete residual assembly, including interior, boundary, symmetry, determinant and section-force terms, is summarised in Fig.~\ref{fig:loss_assembly}.
	
	\section{Training protocol and implementation details}
	\label{sec:training}
	
	\subsection{Geometry banks and collocation sampling}
	
	The final model is trained on a bank of 256 geometries sampled from Eq.~\eqref{eq:geometry_ranges}. The bank is enriched with additional narrow-gauge specimens to improve coverage of the most challenging stress-gradient cases. The present training configuration is fixed-material and fixed-load: the material parameters are those given in Section~\ref{sec:mechanics}, and the imposed displacement is fixed at $\bar u=1~\mathrm{mm}$. Thus the reported model is a geometry-conditioned operator over the DogBone shape family, not a multi-load or material-parameter operator. A separate validation bank of 24 geometries is used for physics diagnostics during training. The physics loss uses 6000 interior collocation points and 2400 boundary points per geometry. Collocation points are resampled during training, which reduces overfitting to a fixed point set and improves residual robustness. The section-force term uses 24 vertical slices with 192 integration points per slice.
	
	\subsection{Network and optimisation parameters}
	
	Table~\ref{tab:training_parameters} summarises the principal architecture and training parameters. The model is trained with Adam, gradient clipping and a learning-rate scheduler. The reported model uses the final physics-based training configuration and is evaluated only after training on independent strict finite-element cases.
	
	\begin{table}[t]
		\centering
		\caption{Principal PI-GINOT architecture and final training parameters.}
		\label{tab:training_parameters}
		\begin{tabular}{ll}
			\toprule
			Quantity & Value \\
			\midrule
			Geometry encoder input & Boundary point cloud $(X,Y)$ \\
			Boundary point-cloud size & 320 \\
			Training geometries & 256 \\
			Validation-bank geometries & 24 \\
			Latent tokens & 32 \\
			Embedding dimension & 64 \\
			Encoder self-attention layers & 3 \\
			Decoder cross-attention layers & 6 \\
			Attention heads & 4 \\
			NeRF positional-encoding maximum degree & 6 \\
			Interior collocation points & 6000 per geometry \\
			Boundary collocation points & 2400 per geometry \\
			Section-force slices & 24 \\
			Section integration points per slice & 192 \\
			Optimiser & Adam \\
			Learning rate & $5\times10^{-5}$ \\
			Gradient clipping & $\|\nabla\|\le 1$ \\
			Principal loss weights & $w_{\rm eq}=150$, $w_{\rm trac}=4$--$60$, $w_N=80$ \\
			\bottomrule
		\end{tabular}
	\end{table}
	
	\subsection{Built-in diagnostics}
	
	Physics-informed operator training can fail in ways that are not evident from the total loss alone. The implementation therefore uses four diagnostic families. First, an initial-state diagnostic checks that the hard-boundary-condition baseline produces finite residuals and positive determinant values before long training. Second, a latent-swap test verifies that the decoder uses the geometry latent: replacing the geometry tokens of one specimen by those of another while keeping query points fixed should change the predicted field. Third, boundary resampling and permutation checks test stability of the geometry encoder with respect to equivalent point-cloud representations of the same boundary. Fourth, the section-force coefficient of variation measures whether the internal axial resultant remains approximately constant along the specimen.
	
	The corresponding training histories, validation-bank residuals and operator-level diagnostic quantities are reported in \ref{app:training_diagnostics}. Qualitative PI-GINOT field predictions for the complete 24-geometry validation bank are provided in \ref{app:validation_bank_fields}.
	
	\section{Strict finite-element validation protocol}
	\label{sec:fem}
	
	A data-free method must be validated without compromising the data-free claim. Finite-element solutions are therefore used only after training and only for assessment. The finite-element reference model uses the same quarter-domain geometry, material parameters, displacement loading and plane-stress constitutive law as PI-GINOT. {\color{black}The calculations are performed in Abaqus/Standard using a custom user element implementing the logarithmic compressible Neo-Hookean model and the local plane-stress closure $P_{33}=0$.}
	
	{\color{black}The user element was implemented specifically for this study so that the finite-element reference model uses the same strain-energy density, stress measure and plane-stress closure as the physics-informed residual. At each quadrature point, the element routine evaluates the in-plane deformation gradient, solves the local plane-stress condition for the out-of-plane stretch, computes the first Piola--Kirchhoff stress and assembles the element residual for Abaqus/Standard. The reference solutions are therefore independent finite-element calculations with matched constitutive assumptions, rather than comparisons against a different built-in material model.}
	
	{\color{black}For each of the eight post-training comparison geometries, a separate Abaqus/Standard analysis was run, and nodal displacements, reaction forces and Gauss-point stress quantities were exported from the output database for subsequent comparison.} The validation set consists of eight independent DogBone geometries spanning the parameter ranges in Eq.~\eqref{eq:geometry_ranges}. PI-GINOT is then evaluated at the same coordinates using the analytic boundary representation of the corresponding geometry. The finite-element data are not used in training, loss weighting or model fitting.
	
\begin{figure}[t]
		\centering
		\maybeincludegraphics[width=0.98\linewidth]{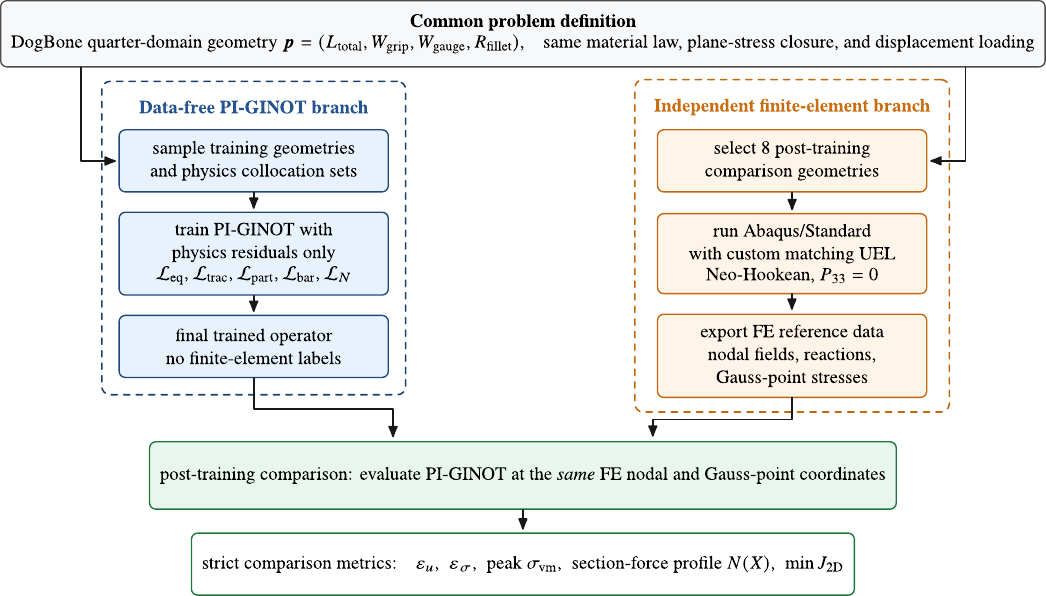}
		\caption{Strict post-training finite-element validation workflow.
			The data-free PI-GINOT branch uses only sampled geometries, collocation points and physics residuals during training. The independent finite-element branch solves eight post-training comparison geometries with a matching Abaqus UEL reference model. After training, PI-GINOT is evaluated at the same finite-element nodal and Gauss-point coordinates, enabling strict comparison of displacement errors, stress errors, peak von Mises stress, section-force profiles and deformation admissibility.}
		\label{fig:validation_protocol}
	\end{figure}

	The complete post-training validation workflow is summarised in Fig.~\ref{fig:validation_protocol}.
	
	\subsection{Validation metrics}
	
	For a common Gauss-point set, the displacement and stress relative errors are
	\begin{equation}
		\varepsilon_u=
		\frac{\|\bu_{\rm NN}-\bu_{\rm FE}\|_{L^2(\Omega)}}
		{\|\bu_{\rm FE}\|_{L^2(\Omega)}},
		\qquad
		\varepsilon_\sigma=
		\frac{\|\bsig_{\rm NN}-\bsig_{\rm FE}\|_{L^2(\Omega)}}
		{\|\bsig_{\rm FE}\|_{L^2(\Omega)}} .
		\label{eq:validation_errors}
	\end{equation}
	Peak von Mises stress is used as a design-oriented scalar stress measure. Internal axial section-force consistency is evaluated from Eq.~\eqref{eq:section_force} using the same vertical sections for the finite-element and PI-GINOT fields. Direct right-edge stress integration is not used as the primary mechanics metric because it is sensitive to boundary quadrature and local post-processing at the constrained edge. The internal section-force profile is more informative for load transfer through the specimen interior.
	
	\section{Results}
	\label{sec:results}
	
	\subsection{Validation geometries}
	
\begin{figure}[t]
		\centering
		\maybeincludegraphics[width=0.98\linewidth]{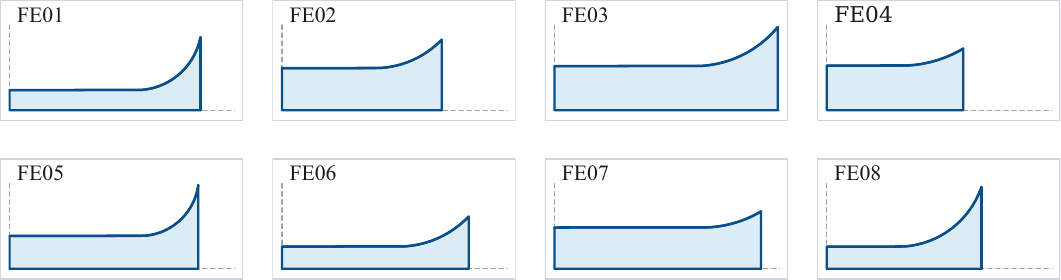}
		\caption{Independent strict-FEM validation geometries. The analysed quarter domains are drawn at common scale for all eight validation cases. Dashed construction lines indicate the two symmetry planes.}
		\label{fig:validation_geometries}
	\end{figure}

The independent post-training finite-element comparison geometries are listed in Table~\ref{tab:validation_geometries} and visualised at common scale in Fig.~\ref{fig:validation_geometries}. The set includes wide-gauge specimens, narrow-gauge specimens and different fillet radii so that both smooth and stress-gradient-dominated cases are tested.
	
	\begin{table}[t]
		\centering
		\caption{Independent strict-FEM validation geometries. All dimensions are in millimetres.}
		\label{tab:validation_geometries}
		\begin{tabular}{lrrrr}
			\toprule
			Case & $L_{\rm total}$ & $W_{\rm grip}$ & $W_{\rm gauge}$ & $R_{\rm fillet}$ \\
			\midrule
			FE01 & 59.405 & 22.639 &  6.239 & 10.134 \\
			FE02 & 49.713 & 21.867 & 13.038 & 14.578 \\
			FE03 & 69.505 & 25.785 & 13.679 & 16.747 \\
			FE04 & 42.451 & 19.109 & 13.821 & 18.857 \\
			FE05 & 58.734 & 25.962 & 10.198 &  9.134 \\
			FE06 & 58.073 & 16.185 &  6.889 & 15.186 \\
			FE07 & 64.318 & 17.780 & 12.803 & 18.036 \\
			FE08 & 48.134 & 25.359 &  6.859 & 13.730 \\
			\bottomrule
		\end{tabular}
	\end{table}
	
	\subsection{Field and scalar error metrics}
	
\begin{figure}[t]
		\centering
		\maybeincludegraphics[width=0.98\linewidth]{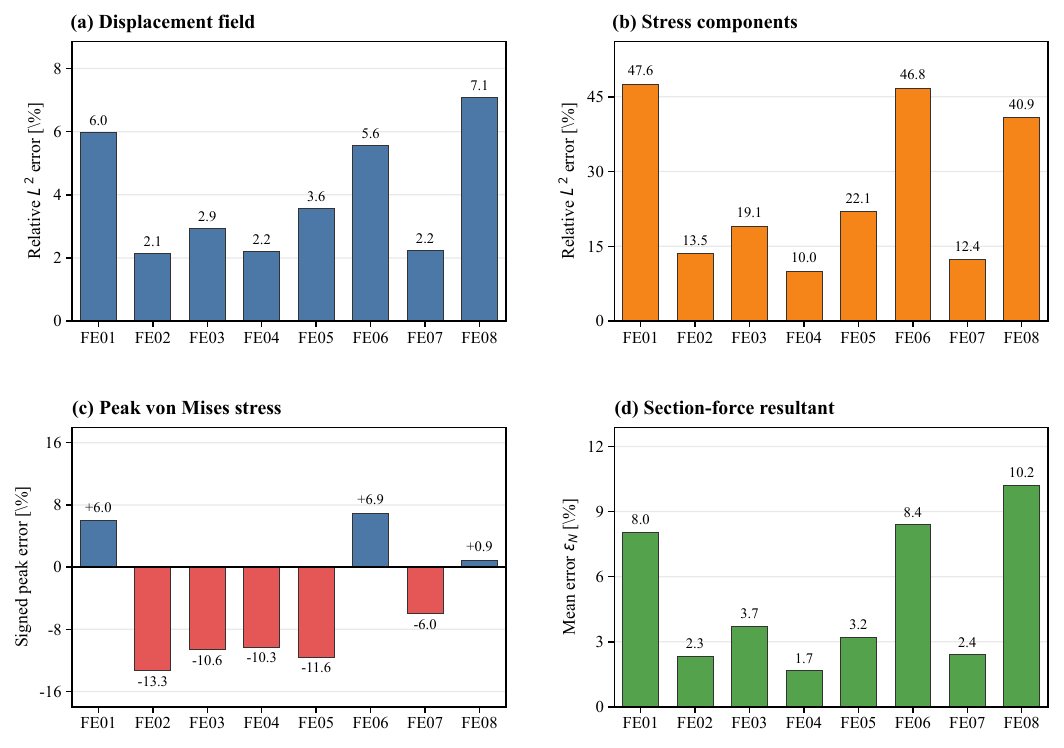}
		\caption{Quantitative strict-FEM validation summary for the final PI-GINOT model.
			Panels (a) and (b) report relative $L^2$ errors for the displacement field and stress components, respectively.
			Panel (c) shows the signed error in the peak von Mises stress, where positive values indicate over-prediction and negative values indicate under-prediction.
			Panel (d) reports the mean relative internal section-force error $\varepsilon_N$.
			The results show consistently low displacement errors, larger component-wise stress errors in the more challenging geometries and bounded global load-transfer errors.}
		\label{fig:validation_summary}
	\end{figure}

	Table~\ref{tab:validation_metrics} reports the main quantitative validation values, while Fig.~\ref{fig:validation_summary} summarises the corresponding error trends across the eight strict-FEM validation cases. Across the eight geometries, the relative displacement error ranges from $2.13\%$ to $7.09\%$. The lowest displacement errors occur in the wider-gauge cases, while the largest displacement error occurs in FE08, a narrow-gauge case with a pronounced stress gradient near the transition region. The peak von Mises stress is recovered within $0.88\%$ to $13.32\%$. The strongest peak-stress agreement occurs in FE08, despite its larger field error, indicating that the model captures the dominant stress concentration magnitude even when the detailed component-wise stress field remains challenging.
	
	\begin{table}[t]
		\centering
		\caption{Strict-FEM validation metrics for the final PI-GINOT model. Displacement and stress errors are relative $L^2$ errors over common Gauss-point locations. The peak von Mises error is signed. The section-force error is the mean absolute relative error of $N(X)$ over internal sections.}
		\label{tab:validation_metrics}
		\begin{tabular}{lrrrrrr}
			\toprule
			Case & $\varepsilon_u$ [\%] & $\varepsilon_\sigma$ [\%] & Peak VM err. [\%] & FE peak VM [MPa] & NN peak VM [MPa] & $\varepsilon_N$ [\%] \\
			\midrule
			FE01 & 5.98 & 47.56 &  5.95 & 31.61 & 33.49 &  8.05 \\
			FE02 & 2.13 & 13.50 & -13.32 & 38.32 & 33.22 &  2.35 \\
			FE03 & 2.95 & 19.10 & -10.63 & 27.41 & 24.49 &  3.71 \\
			FE04 & 2.20 & 10.04 & -10.30 & 42.76 & 38.36 &  1.66 \\
			FE05 & 3.58 & 22.05 & -11.57 & 33.71 & 29.81 &  3.20 \\
			FE06 & 5.56 & 46.80 &  6.91 & 31.25 & 33.41 &  8.39 \\
			FE07 & 2.25 & 12.42 & -6.00 & 27.92 & 26.24 &  2.41 \\
			FE08 & 7.09 & 40.91 &  0.88 & 40.69 & 41.05 & 10.23 \\
			\bottomrule
		\end{tabular}
	\end{table}
	
	The stress-component errors are larger than the displacement errors, ranging from $10.0\%$ to $47.6\%$ and are particularly large for FE01, FE06 and FE08. This behaviour is expected because stresses are derived from displacement gradients and are therefore more sensitive to local errors near the curved gauge-fillet transition. It is also mechanically meaningful that the narrow-gauge cases are the most difficult: they combine smaller load-carrying width with sharper local stress variations.
	
	\subsection{Representative field comparisons}
	
\begin{figure}[t]
		\centering
		\maybeincludegraphics[width=0.98\linewidth]{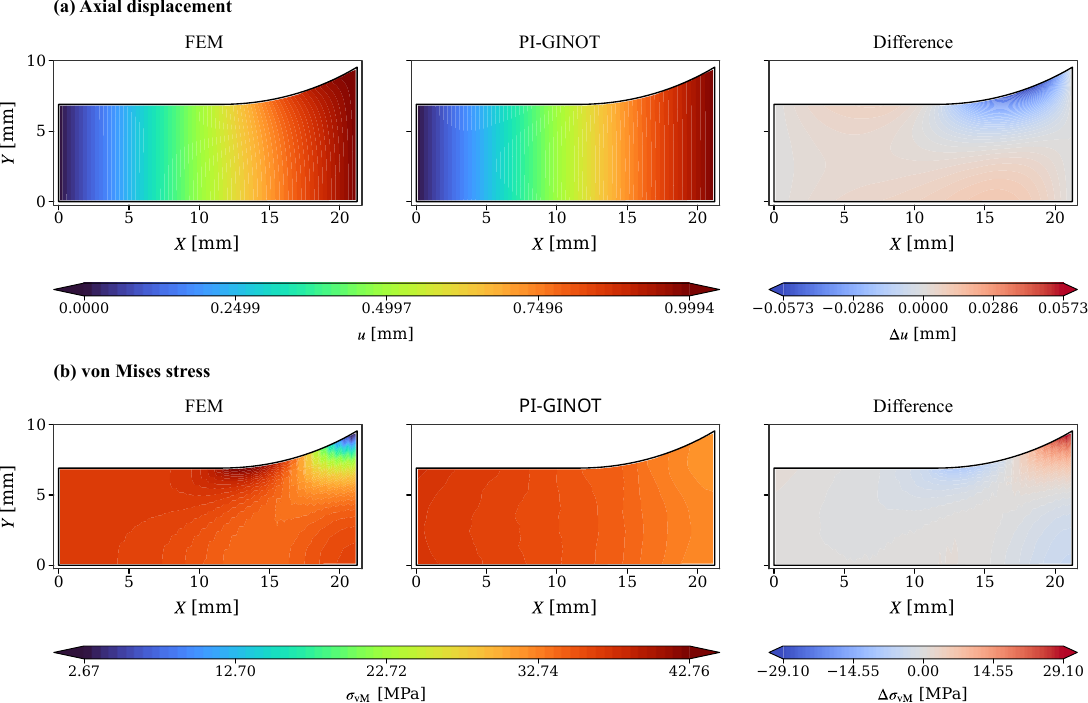}
		\caption{Representative strong validation case FE04. Strict finite-element reference fields are compared with the final PI-GINOT prediction at common Gauss-point coordinates. Panel (a) shows the axial displacement field $u$, and panel (b) shows the von Mises stress field $\sigma_{\mathrm{vM}}$. The third column shows the signed difference between PI-GINOT and FEM fields.}
		\label{fig:field_strong}
	\end{figure}

\begin{figure}[t]
		\centering
		\maybeincludegraphics[width=0.98\linewidth]{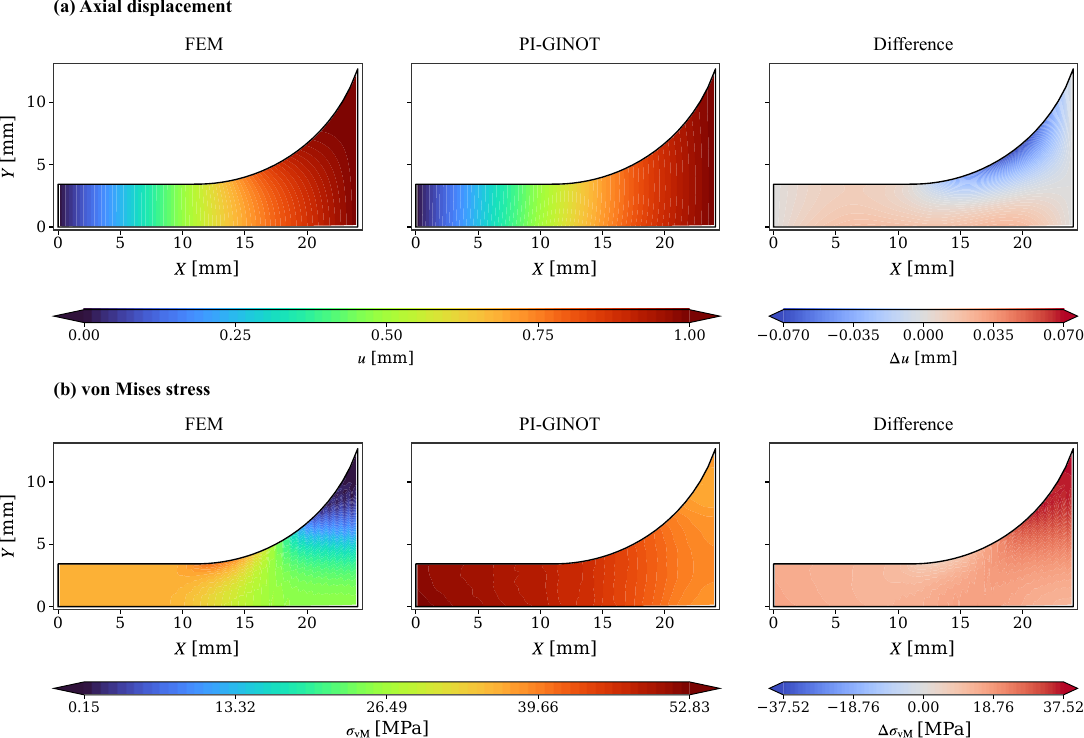}
		\caption{Challenging narrow-gauge validation case FE08. Strict finite-element reference fields are compared with the final PI-GINOT prediction at common Gauss-point coordinates. Panel (a) shows the axial displacement field $u$, and panel (b) shows the von Mises stress field $\sigma_{\mathrm{vM}}$. The third column shows the signed difference between PI-GINOT and FEM fields.}
		\label{fig:field_challenging}
	\end{figure}

	Representative field comparisons are included for one strong case and one challenging case. 
	FE04 is selected as the strong case because it combines low displacement error, low component-wise stress error and the lowest section-force error; its field comparison is shown in Fig.~\ref{fig:field_strong}. 
	FE08 is selected as the challenging case because it has the largest displacement and section-force errors while retaining excellent peak von Mises agreement; its field comparison is shown in Fig.~\ref{fig:field_challenging}. 
	This pairing gives an honest view of both the capability and the limitation of the final model.

	\subsection{Internal section-force consistency}
	
\begin{figure}[t]
		\centering
		\includegraphics[width=0.98\linewidth]{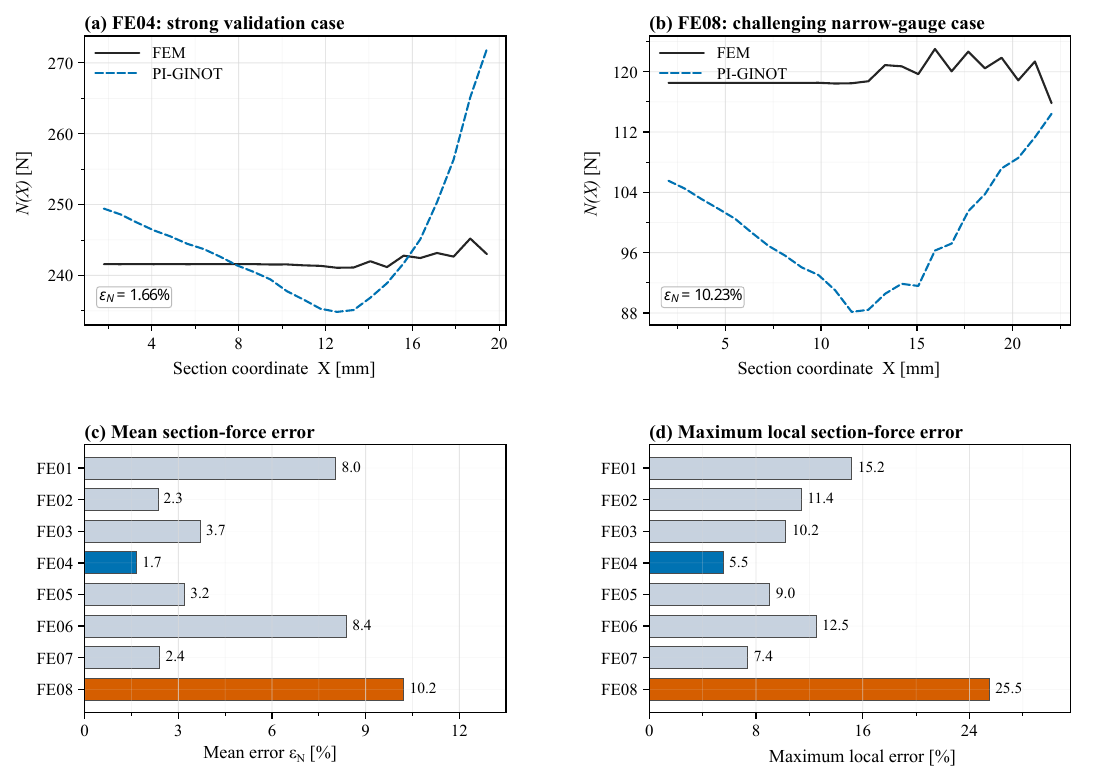}
		\caption{Internal axial section-force consistency for the strict-FEM validation cases.
			Panels (a) and (b) compare the section-force profiles $N(X)$ predicted by PI-GINOT with the corresponding finite-element reference profiles for the representative strong case FE04 and the challenging narrow-gauge case FE08.
			Panels (c) and (d) summarise the mean section-force error and the maximum local section-force error over all eight strict-FEM validation geometries.
			The results show that the final PI-GINOT model reproduces the global axial load-transfer behaviour, while the largest local deviations occur for the narrow-gauge FE08 case.}
		\label{fig:section_force}
	\end{figure}

	Table~\ref{tab:section_force} reports the section-force statistics, while Fig.~\ref{fig:section_force} visualises the corresponding internal axial load-transfer behaviour. 
	To further assess whether the predicted stress fields preserve global load transfer, the internal section-force profiles $N(X)$ are compared between PI-GINOT and the strict-FEM references. 
	The section force is obtained by integrating the first Piola--Kirchhoff stress component over transverse sections. 
	This diagnostic is complementary to pointwise stress errors because it measures whether local stress predictions remain mechanically consistent at the resultant level.
	
	\begin{table}[t]
		\centering
		\caption{Internal section-force consistency. $\mathrm{CV}_{\rm FE}$ and $\mathrm{CV}_{\rm NN}$ are coefficients of variation of $N(X)$. $\varepsilon_N$ is the mean absolute relative section-force error.}
		\label{tab:section_force}
		\begin{tabular}{lrrrrr}
			\toprule
			Case & $\mathrm{CV}_{\rm FE}$ [\%] & $\mathrm{CV}_{\rm NN}$ [\%] & $\overline{N}_{\rm FE}$ [N] & $\overline{N}_{\rm NN}$ [N] & $\varepsilon_N$ [\%] \\
			\midrule
			FE01 & 1.03 &  6.19 &  82.39 &  77.13 &  8.05 \\
			FE02 & 0.49 &  4.24 & 198.03 & 199.77 &  2.35 \\
			FE03 & 0.56 &  4.57 & 150.22 & 148.07 &  3.71 \\
			FE04 & 0.25 &  2.00 & 241.73 & 244.95 &  1.66 \\
			FE05 & 0.65 &  3.98 & 131.34 & 129.72 &  3.20 \\
			FE06 & 0.75 &  4.09 &  92.94 &  85.20 &  8.39 \\
			FE07 & 0.49 &  2.65 & 147.68 & 144.65 &  2.41 \\
			FE08 & 0.93 & 11.86 & 118.82 & 119.30 & 10.23 \\
			\bottomrule
		\end{tabular}
	\end{table}
	
	The finite-element references exhibit nearly constant section-force profiles, as expected for the displacement-controlled equilibrium problem. The PI-GINOT profiles follow this behaviour closely for the strong validation case FE04, which has the smallest mean section-force error. The narrow-gauge FE08 case shows the largest local profile deviations, although its mean resultant remains close to the finite-element value. This confirms that the main limitation is local stress-gradient resolution rather than global load-transfer prediction.

	\section{Discussion}
	\label{sec:discussion}
	
	\subsection{Scope of the evidence}
	
	The evidence in this paper should be interpreted within the controlled scope of the benchmark. The final PI-GINOT model is trained and assessed for a fixed material, a fixed imposed displacement, a plane-stress compressible Neo-Hookean formulation and a smooth single-topology DogBone geometry family. The comparison cases vary the DogBone geometry within this family, but they do not test holes, notches, disconnected boundaries, contact, topology changes, heterogeneous materials or arbitrary loading paths. This scope is intentional. It isolates the question of whether a boundary point cloud can condition a reusable, data-free neural operator for finite-strain solid mechanics when the governing equations, boundary conditions and constitutive law are enforced directly.
	
	Within this bounded setting, the results are mechanically meaningful. The model is not supervised by finite-element displacement or stress labels, yet it predicts admissible displacement fields, captures peak stress levels with useful accuracy and reproduces internal axial load transfer with bounded section-wise deviations. The finite-element simulations therefore play a validation role rather than a training role. This distinction is central to the claim: the method is data-free with respect to FEM field labels, but it is still assessed against an independent strict-FEM reference solution.
	
	\subsection{Metric non-equivalence and error hierarchy}
	
	The validation results show a clear hierarchy of difficulty. Displacement errors are the smallest, peak von Mises stress errors are moderate and component-wise stress-field errors are the largest. This hierarchy is expected in finite-strain neural mechanics. The displacement field is the primary neural output and is constrained directly by the hard boundary-condition layer. Stresses are derived quantities: they depend on spatial derivatives of the displacement field, the deformation gradient, the plane-stress closure and the nonlinear hyperelastic constitutive law. Small displacement-gradient errors can therefore be amplified in stress components, especially near high-curvature boundaries and stress concentrations.
	
	Different metrics also measure different physical properties. A low displacement error does not imply a low component-wise stress error. A low peak von Mises error does not imply that the entire stress field is accurate point by point. Similarly, agreement of the average axial force does not guarantee that the internal section-force profile is correct at every section. FE08 illustrates this distinction particularly well. Its peak von Mises error is less than $1\%$, and its mean finite-element and PI-GINOT axial resultants are almost identical. However, its stress-field error and section-wise force-profile error are among the largest in the comparison set. This behaviour is not contradictory; it means that the model recovers some global and scalar quantities well while still misrepresenting parts of the local stress-gradient structure.
	
	This metric non-equivalence is important for engineering interpretation. If the intended use is rapid screening of geometry families, stiffness trends or global load transfer, the present accuracy may already be useful. If the intended use is fatigue initiation, fracture initiation or certification-level local stress assessment, the component-wise stress errors in narrow-gauge cases remain too high. The model should therefore be viewed as a promising geometry-conditioned mechanics surrogate, not as a replacement for refined finite-element verification of local stress hot spots.
	
	\subsection{Role of hard boundary-condition enforcement}
	
	The hard boundary-condition layer is one of the most important modelling choices in the present formulation. It enforces $u(0,Y)=0$, $u(L_{\rm half},Y)=\bar u$ and $v(X,0)=0$ exactly for any values of the raw decoder outputs. Consequently, the network does not need to learn the essential constraints through a weighted penalty that competes with equilibrium and traction terms. This is especially useful in finite-strain mechanics because inaccurate displacement constraints can corrupt the deformation gradient, stress field and apparent reaction response.
	
	Exact enforcement does not by itself guarantee stress accuracy, but it provides a kinematically reliable foundation for the physics-informed residuals. The remaining errors are therefore more clearly attributable to local derivative resolution, stress-gradient representation and geometry conditioning, rather than to drift in the essential boundary conditions. The cost of this advantage is reduced generality: the admissible layer is tailored to the present boundary-condition topology. Extending the same idea to more complex domains will require automated distance functions, constructive boundary descriptors or learned admissibility maps that preserve exactness without hand-crafting a new layer for every support and loading configuration.
	
	\subsection{Why section-force consistency matters}
	
	The internal section-force diagnostic is more than an additional scalar error. The resultant $N(s)$, defined in Eq.~\eqref{eq:section_force} for a vertical section at $X=s$, should remain nearly constant along the specimen in the absence of body forces, up to numerical and interpolation errors. This is a global consequence of equilibrium. A model can have visually plausible fields and even acceptable pointwise residuals on sampled collocation points while still drifting in integrated load transfer. Conversely, a model may produce imperfect local stress components while preserving the main axial load path.
	
	The strict-FEM references show very small coefficients of variation in $N(s)$, confirming that they provide a reliable baseline for internal load transfer. The final PI-GINOT model reproduces this behaviour best in the broad-gauge cases, with mean section-force errors between approximately $1.7\%$ and $3.7\%$ for the strongest cases. The narrow-gauge cases remain more difficult because the same axial load is transmitted through a smaller and more rapidly varying cross-section, and because the stress field near the gauge-fillet transition is more sensitive to derivative errors. The FE08 result should be read in this profile-based sense: the mean resultant level is nearly correct, but the spatial variation of $N(s)$ is less accurate. This is precisely why section-force consistency is valuable; it reveals errors that would be hidden by a single averaged force.
	
	\subsection{Stress concentration and geometry generalisation}
	
	The DogBone benchmark is simple in topology but nontrivial in geometry dependence. Changing $W_{\rm gauge}$ and $R_{\rm fillet}$ changes both the nominal stress level and the stress concentration near the transition region. A geometry-independent PINN-like surrogate could fit an average deformation trend but would not know how the local stress concentration should shift and scale with the boundary shape. PI-GINOT addresses this by encoding the boundary point cloud into latent geometry tokens and conditioning query-point predictions on those tokens through cross-attention and modulation.
	
	The comparison cases indicate that this representation captures the dominant geometry dependence of the displacement field and the peak-stress magnitude. The remaining errors show where the representation is still incomplete. Narrow gauges and sharp gauge-to-grip transitions demand more accurate local derivatives than the current smooth neural decoder provides. Future improvements should therefore target local stress-gradient resolution rather than only global loss reduction. Promising directions include curvature-biased boundary sampling, adaptive collocation near the fillet, derivative-aware regularisation, local stress-enrichment heads and multi-scale query features.
	
	\subsection{Value of strict finite-element validation}
	
	Physics-informed loss convergence is not sufficient evidence of predictive accuracy in mechanics. Residuals depend on point sampling, residual scaling, loss weights and the numerical distribution of collocation points. A trained model can reduce its loss while still producing inaccurate stresses in regions that are under-sampled or difficult to differentiate. For this reason, the present paper separates training from validation. {\color{black}The finite-element reference solutions are produced independently in Abaqus/Standard using the custom user element described in Section~\ref{sec:fem}, with the same reference geometry, material law, plane-stress closure and imposed displacement.}
	
	This validation design strengthens the interpretation of the results. The comparison quantifies not only whether the neural fields look smooth, but how they compare against a conventional mechanics discretisation. It also reveals where the method is reliable and where it is not. The displacement and peak-stress results support PI-GINOT as a reusable surrogate for the studied DogBone family. The stress-component and section-force profile errors identify the current limits of local stress and load-path fidelity. This balanced evidence is preferable to claiming broad predictive accuracy from residual minimisation alone.
	
	\subsection{Implications, limitations and future work}
	
	The results suggest that data-free geometry-conditioned neural operators can occupy a useful middle ground between single-instance PINNs and fully supervised neural operators. A standard PINN can avoid labelled data but usually lacks geometry reusability. A supervised neural operator can provide fast multi-query inference but requires a database of high-fidelity solutions. PI-GINOT combines a reusable operator structure with physics-informed training, reducing dependence on finite-element snapshots while retaining a mechanics-based objective.
	
	The present study is nevertheless a first controlled step. It does not yet demonstrate arbitrary-domain generalisation, multi-load response, material-parameter conditioning or topology changes. It also does not include a complete ablation study separating the contributions of the hard boundary-condition layer, determinant barrier, traction residuals and section-force term. Computational efficiency should be quantified in a dedicated timing study, including geometry encoding time, decoding time per query point, stress post-processing time and comparison with finite-element solve time. Despite these limitations, the present results establish a coherent proof of concept: a boundary point cloud can condition a reusable neural operator for finite-strain solid mechanics, exact boundary-condition enforcement improves kinematic reliability, and internal section-force consistency provides a physically interpretable global check for data-free operator learning.
	
	\section{Conclusions}
	\label{sec:conclusion}
	
	This paper introduced PI-GINOT, a physics-informed geometry-informed neural operator transformer for data-free finite-strain hyperelasticity on parametric DogBone specimens. The method represents each geometry by a boundary point cloud, encodes it into latent geometry tokens and decodes displacement fields at arbitrary query points through cross-attention. Essential displacement boundary conditions are imposed exactly by a hard boundary-condition layer, and stresses are obtained by automatic differentiation and a plane-stress compressible Neo-Hookean material model.
	
	The model is trained without finite-element displacement or stress labels. The objective combines reference-configuration equilibrium, traction-free boundary residuals, partial traction components on symmetry boundaries, determinant-barrier regularisation and internal axial section-force consistency. Independent strict-FEM simulations are used only after training for validation. Across eight independent comparison geometries, the final PI-GINOT model achieves displacement errors of $2.1\%$--$7.1\%$, peak von Mises stress errors within $0.9\%$--$13.3\%$ and mean section-force errors not exceeding $10.3\%$. The largest remaining errors occur in component-wise stress fields, which range from $10.0\%$ to $47.6\%$ and are concentrated in narrow-gauge specimens, where local stress gradients are strongest.
	
	The results show that a data-free, geometry-conditioned operator can provide mechanically meaningful predictions for the studied family of nonlinear solid-mechanics boundary-value problems. The study also highlights the importance of exact boundary-condition enforcement, metric-aware interpretation, internal load-transfer diagnostics and independent finite-element validation for evaluating physics-informed neural operators in mechanics.
	

	
	\clearpage
	\appendix
	
	\renewcommand{\thefigure}{\thesection.\arabic{figure}}
	\setcounter{figure}{0}
	
	\section{Training diagnostics}
	\label{app:training_diagnostics}
	
\begin{figure}[p]
		\centering
		\includegraphics[width=0.98\textwidth]{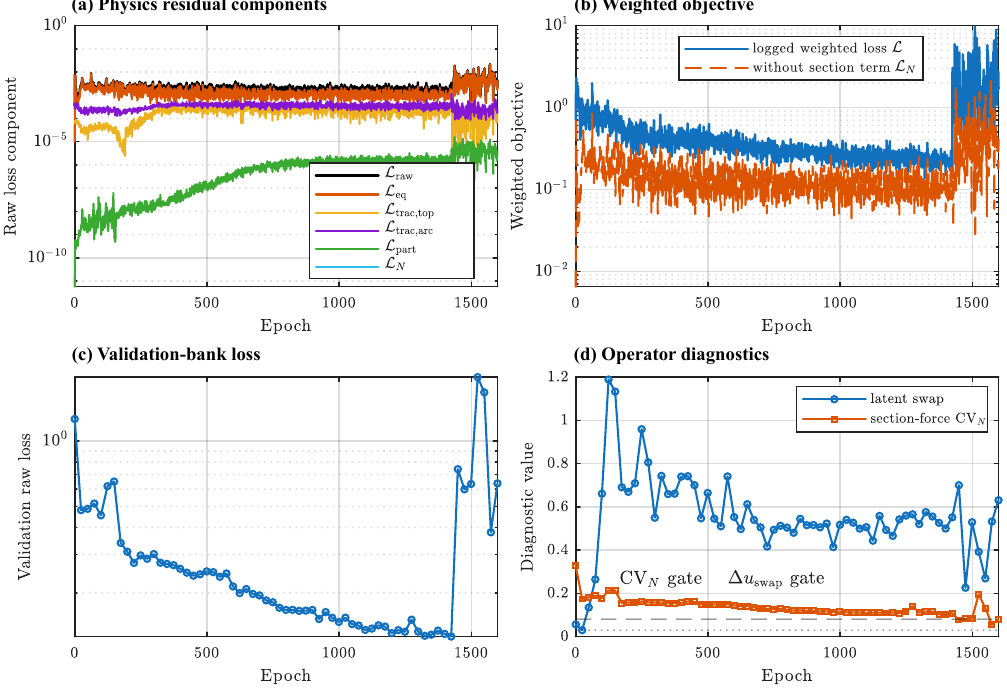}
		\caption{Training diagnostics for the final PI-GINOT optimisation.
			The figure reports raw physics residual components, the weighted objective, the validation-bank residual and two operator-level diagnostics.
			The validation-bank residual decreases over most of the optimisation history, while late-stage fluctuations are observed after approximately 1450 epochs.
			These quantities are used as physics-informed training diagnostics only; no finite-element displacement or stress labels are used during training, and quantitative predictive accuracy is assessed independently using the strict-FEM validation cases reported in the main text.}
		\label{fig:appendix_training_diagnostics}
	\end{figure}

	Figure~\ref{fig:appendix_training_diagnostics} reports the diagnostic quantities monitored during the final physics-informed PI-GINOT optimisation. 
	These quantities include raw residual components, the weighted objective, the validation-bank residual and two operator-level diagnostics. 
	They are used to monitor optimisation stability and geometry sensitivity, but they are not finite-element error measures. 
	The quantitative predictive accuracy of the final model is therefore assessed independently using the strict-FEM validation cases reported in the main text.
	
	\section{Validation-bank field predictions}
	\label{app:validation_bank_fields}
	
	\setcounter{figure}{0}
	
\begin{figure}[p]
		\centering
		\includegraphics[width=0.98\textwidth]{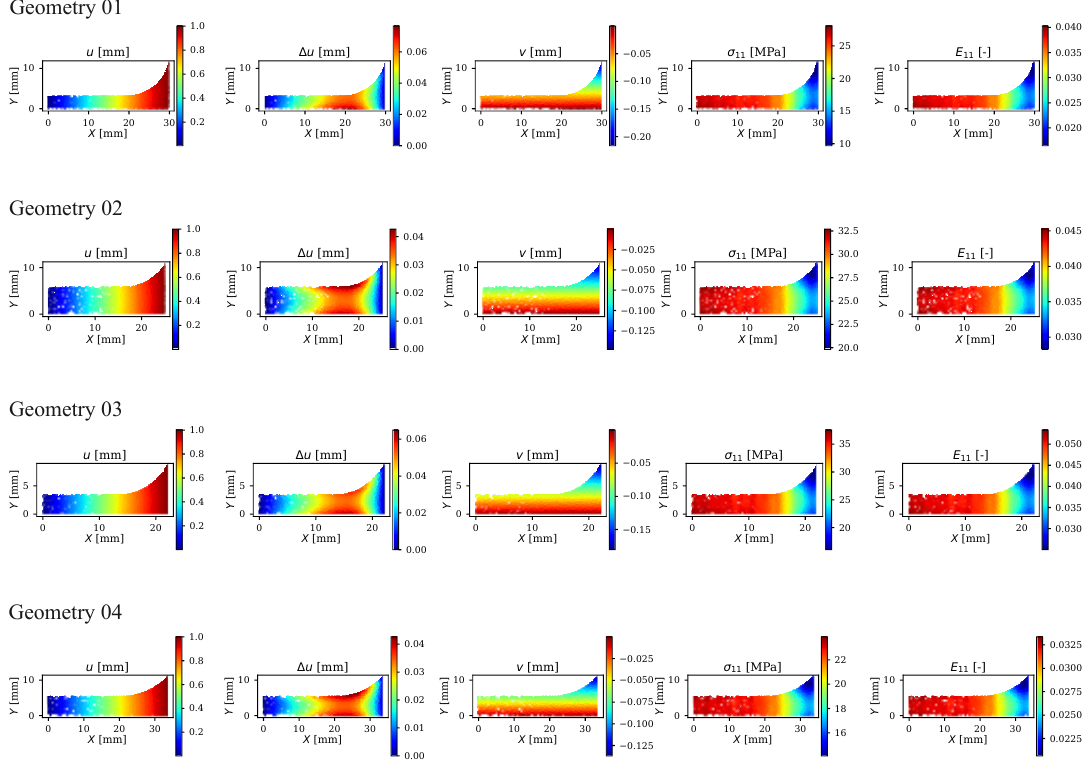}
		\caption{PI-GINOT validation-bank field predictions for validation geometries 01--04.}
		\label{fig:appendix_validation_bank_fields_1}
	\end{figure}

\begin{figure}[p]
		\centering
		\includegraphics[width=0.98\textwidth]{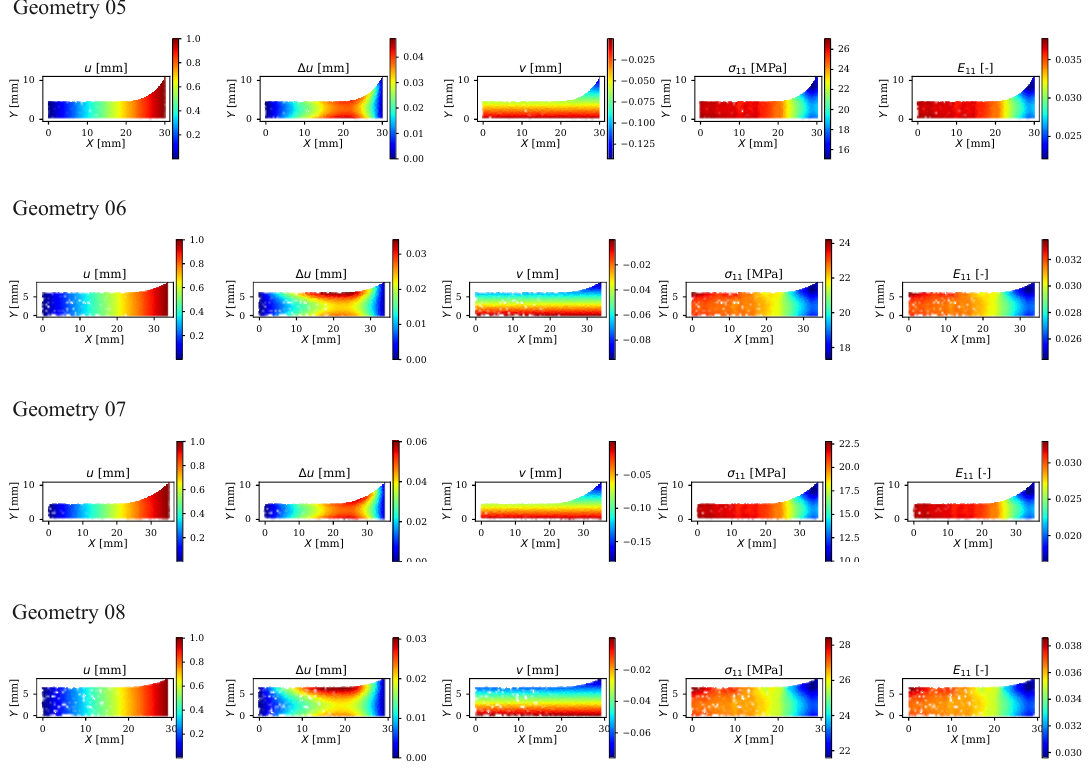}
		\caption{PI-GINOT validation-bank field predictions for validation geometries 05--08.}
		\label{fig:appendix_validation_bank_fields_2}
	\end{figure}

\begin{figure}[p]
		\centering
		\includegraphics[width=0.98\textwidth]{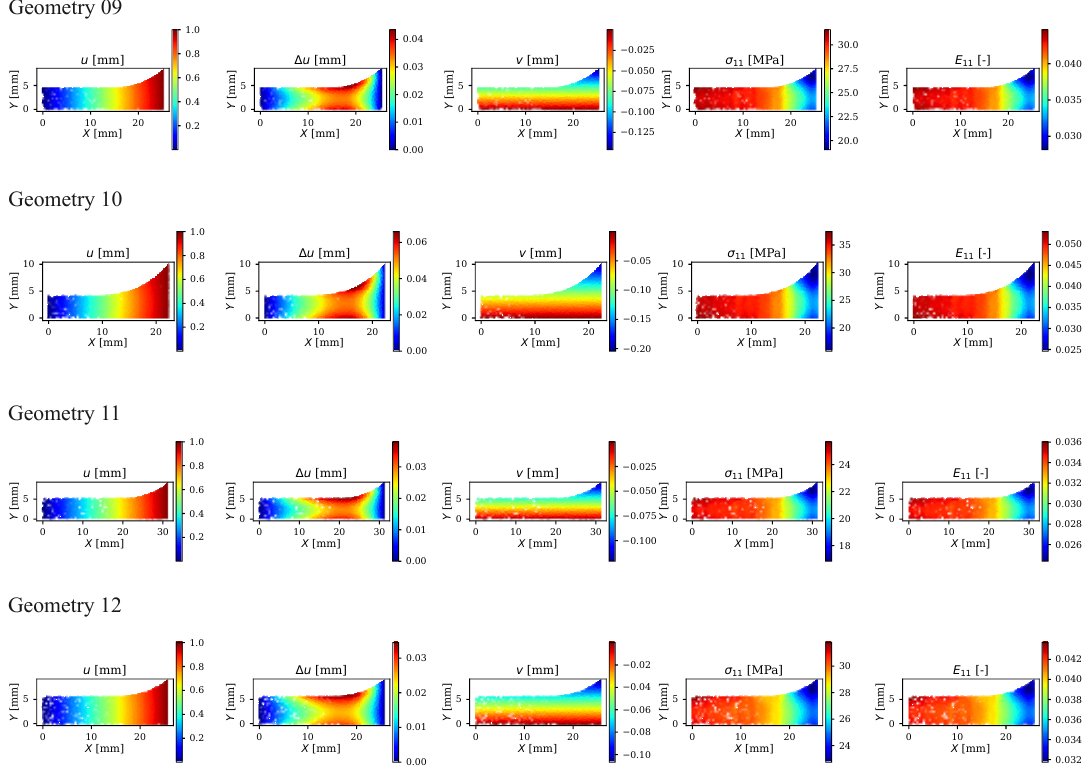}
		\caption{PI-GINOT validation-bank field predictions for validation geometries 09--12.}
		\label{fig:appendix_validation_bank_fields_3}
	\end{figure}

\begin{figure}[p]
		\centering
		\includegraphics[width=0.98\textwidth]{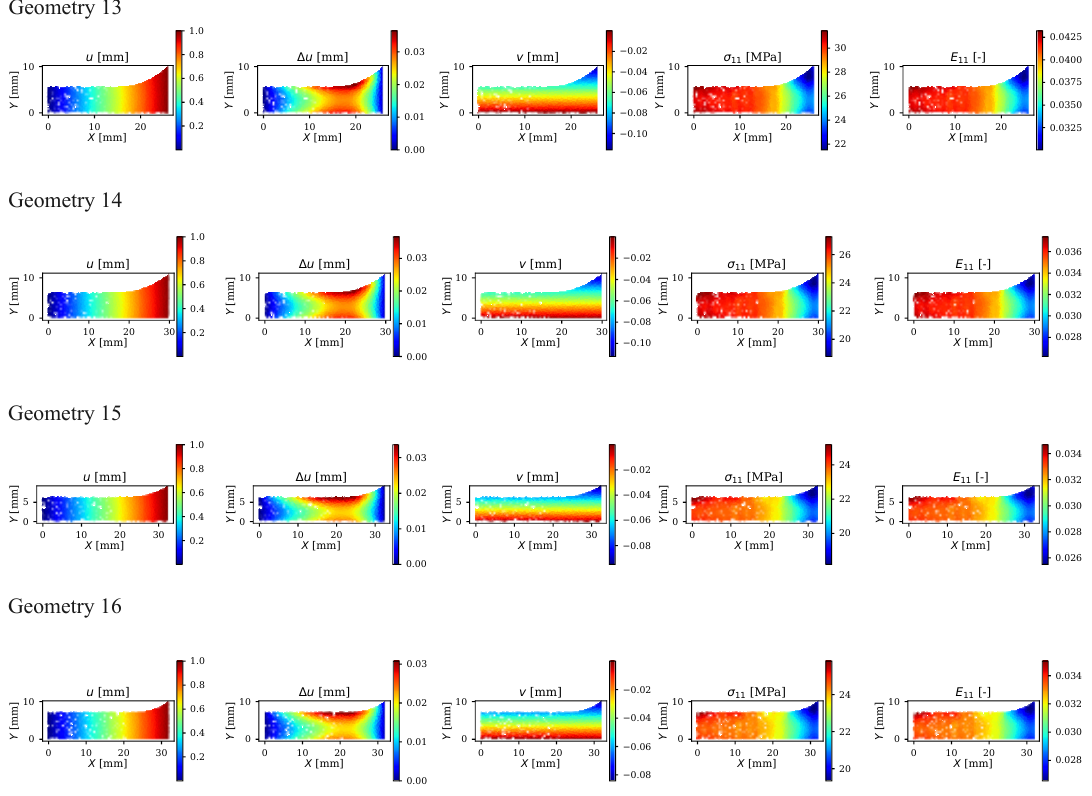}
		\caption{PI-GINOT validation-bank field predictions for validation geometries 13--16.}
		\label{fig:appendix_validation_bank_fields_4}
	\end{figure}

\begin{figure}[p]
		\centering
		\includegraphics[width=0.98\textwidth]{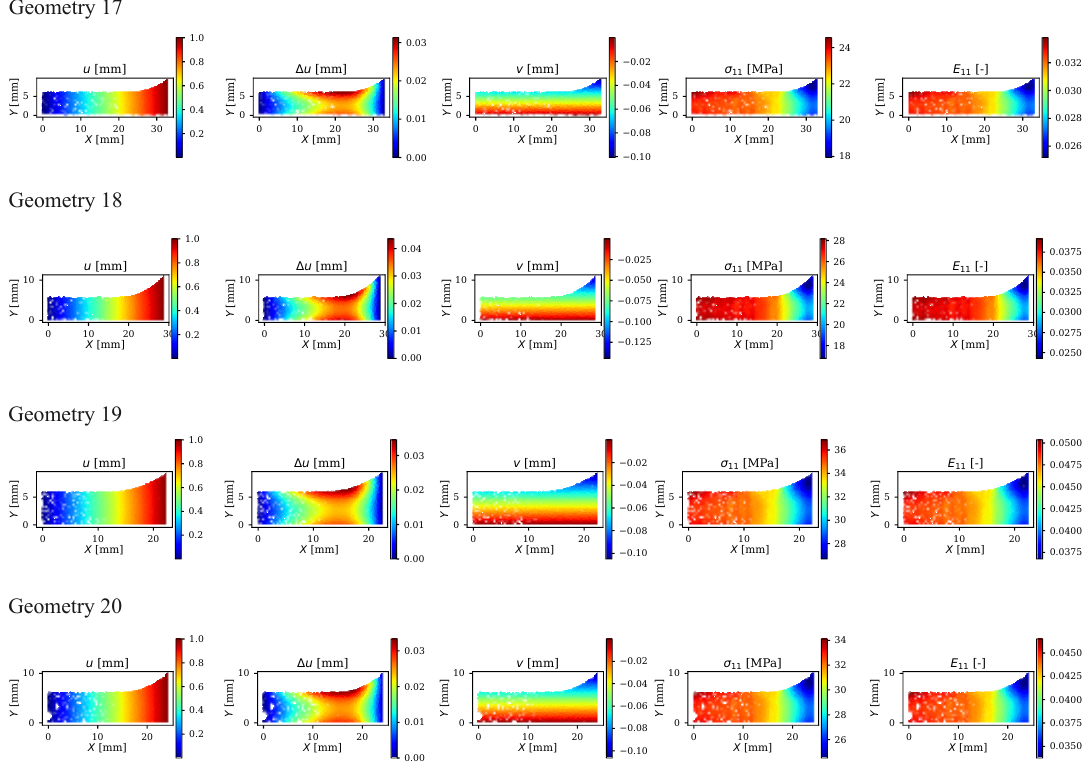}
		\caption{PI-GINOT validation-bank field predictions for validation geometries 17--20.}
		\label{fig:appendix_validation_bank_fields_5}
	\end{figure}

\begin{figure}[p]
		\centering
		\includegraphics[width=0.98\textwidth]{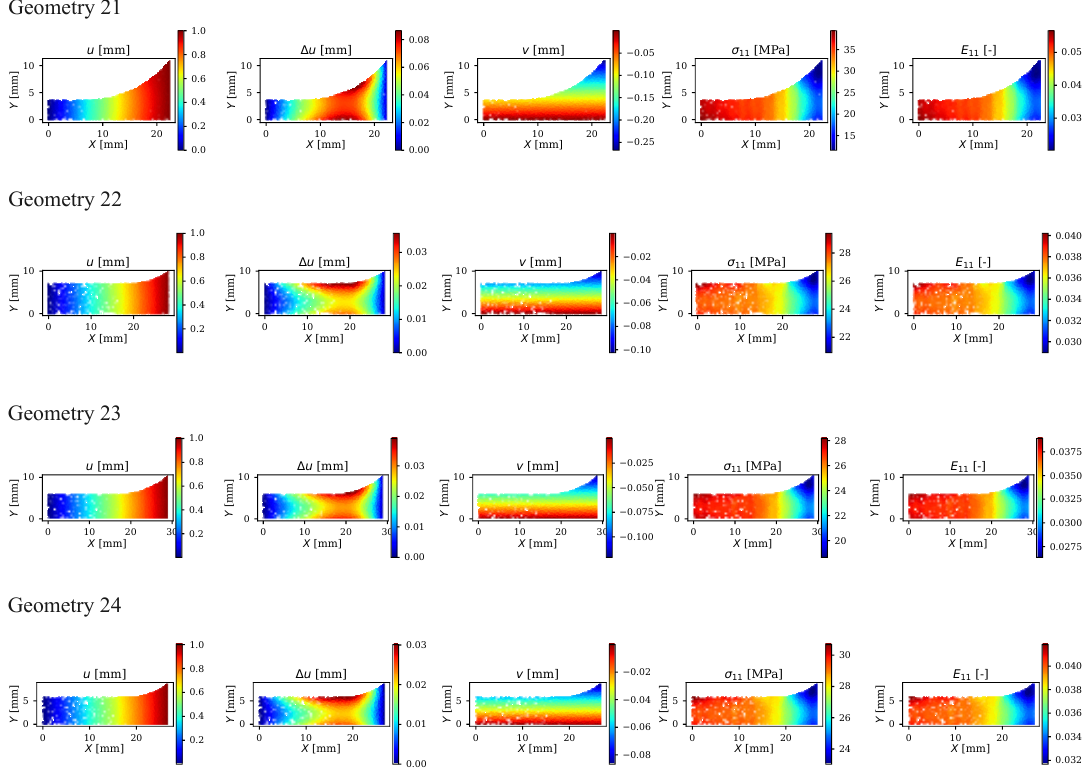}
		\caption{PI-GINOT validation-bank field predictions for validation geometries 21--24.}
		\label{fig:appendix_validation_bank_fields_6}
	\end{figure}

	Figure~\ref{fig:appendix_validation_bank_fields_1}--\ref{fig:appendix_validation_bank_fields_6} show qualitative PI-GINOT field predictions for the complete 24-geometry validation bank used for physics-informed monitoring.
	The plots include the axial displacement $u$, the deviation from the imposed linear displacement mode $\Delta u$, the transverse displacement $v$, the stress component $\sigma_{11}$ and the Green--Lagrange strain component $E_{11}$.
	These fields are not finite-element supervision data; they are post-training PI-GINOT evaluations used to assess qualitative smoothness and geometry-dependent response across the validation-bank parameter range.
	
	\clearpage
	
	\section*{Data and code availability}
	{\color{black} The PI-GINOT implementation will be made available in a public repository upon publication.}
	
	\section*{Declaration of competing interest}
	The authors declare that they have no known competing financial interests or personal relationships that could have appeared to influence the work reported in this paper.
	
	\section*{Declaration of generative AI and AI-assisted technologies in the manuscript preparation process}
	During the preparation of this work, the author used AI-assisted tools for language editing and grammar checking. After using these tools, the author reviewed and edited the content as needed and takes full responsibility for the content of the publication.


\end{document}